\documentclass[reprint,onecolumn,superscriptaddress,showpacs,nofootinbib,notitlepage]{revtex4-1}

\usepackage{graphicx}
\usepackage{latexsym,amsmath,amssymb,lmodern,float,url}
\usepackage{natbib}
\usepackage{color}
\usepackage{microtype}
\usepackage{slashed}
\usepackage{multirow}

\newcommand{\be}{\begin{equation}}
\newcommand{\ee}{\end{equation}}
\newcommand{\bea}{\begin{eqnarray}}
\newcommand{\eea}{\end{eqnarray}}

\newcommand{\nn}{\nonumber}
\newcommand{\ba}{\begin{array}}
\newcommand{\ea}{\end{array}}

\newcommand{\im}{\mathop{\rm{Im}}}

\usepackage[colorlinks=true,backref=false, linktocpage=true,
citecolor=blue,urlcolor=blue,linkcolor=blue,pdfpagemode=UseOutlines]{hyperref}

\hypersetup{%
  bookmarksnumbered=true,
  pdftitle = {},
  pdfsubject = {},
  pdfauthor = {},
  pdfkeywords = {}
}

\def\Re {\operatorname{{Re}}}

\def\Tr {\operatorname{{Tr}}}

\renewcommand{\d}{\mathrm d}

\begin{document}
\title{Finite-Density Monte Carlo Calculations on Sign-Optimized Manifolds}

\author{Andrei Alexandru}
\email{aalexan@gwu.edu}
\affiliation{Department of Physics, The George Washington University, Washington, D.C. 20052, USA}
\affiliation{Department of Physics, University of Maryland, College Park, MD 20742, USA}
\affiliation{Albert Einstein Center for Fundamental Physics, Institute for Theoretical Physics, University of Bern, Sidlerstrasse 5, CH-3012 Bern, Switzerland}
\author{Paulo F. Bedaque}
\email{bedaque@umd.edu}
\affiliation{Department of Physics, University of Maryland, College Park, MD 20742, USA}
\author{Henry Lamm}
\email{hlamm@umd.edu}
\affiliation{Department of Physics, University of Maryland, College Park, MD 20742, USA}
\author{Scott Lawrence}
\email{srl@umd.edu}
\affiliation{Department of Physics, University of Maryland, College Park, MD 20742, USA}
\date{\today}

\newcommand{\Seff}{S_{\text{eff}}}
\begin{abstract}
We present a general technique for addressing sign problems that arise in Monte Carlo simulations of field theories. This method deforms the domain of the path integral to a manifold in complex field space that maximizes the average sign (therefore reducing the sign problem) within a parameterized family of manifolds. We presents results for the $1+1$ dimensional Thirring model with Wilson fermions on lattice sizes up to $40\times 10$.  This method reaches higher $\mu$ then previous techniques while substantially decreasing the computational time required.
\end{abstract}

\maketitle

\section{Introduction}

Monte Carlo methods are critical to the study of field-theoretical and many-body systems. In particular, they are the only general-purpose approach to address  strongly interacting field theories. 
The basic idea of all Monte Carlo methods is simple: observables are formulated as path integrals  which, on a discretized spacetime, become high dimensional integrals. Those are then estimated  stochastically  by importance sampling. Importance sampling relies
on interpreting part of the integrand (typically the exponential of the action) as a probability, which makes sense only if this term is real and non-negative.
Unfortunately, many theories, even when formulated in imaginary time (Euclidean space), have a negative or even complex integrand. This  so-called ``sign problem" is a major roadblock to the understanding of some of the most important systems in physics.  Many systems at finite density (including QCD at finite baryon density) and non-relativistic systems lacking some special symmetry between fermion species (as in the Hubbard model away from half-filling or on non bi-partite lattices) suffer from sign problems. Also, some real time observables in thermal equilibrium as well as truly non-equilibrium phenomena are not amenable to imaginary time calculations and have a particularly severe sign problem that renders most Monte Carlo methods a non-starter. 
A simple, albeit not very effective, way of dealing with the sign problem is to choose a manifestly positive part of the integrand as the statistical weight while moving the part with the fluctuating sign/phase to the observable to be measured. This ``reweighting" is effective to the extent that the average sign, that is, the average of the fluctuating sign on the ensemble defined by the positive measure, is not too small. However, in theories with sign problems, the average sign typically decreases exponentially with the volume and the inverse temperature of the system. 
Many techniques have been proposed in the past to ameliorate the sign problem. Among them are the complex Langevin method~\cite{Aarts:2008rr}, the density of states method~\cite{Langfeld:2016mct}, canonical methods~\cite{Alexandru:2005ix,deForcrand:2006ec}, reweighting methods~\cite{Fodor:2001au}, series expansion in the chemical potential~\cite{Allton:2002zi}, fermion bags~\cite{Chandrasekharan:2013rpa}, and analytic continuation from imaginary chemical potentials~\cite{deForcrand:2006pv}. Each one has its successes and pitfalls. It is fair to say, however, that the sign problems of field theories remain largely unsolved.


More recently, the ``thimble" method was proposed~\cite{Cristoforetti:2012su,Cristoforetti:2013qaa}. The main idea is to complexify the domain of the path integral. Instead of integrating over real values of the fields, one deforms the manifold of integration from $\mathbb R^N \subset \mathbb C^N$ to some other $N$-dimensional manifold, $\mathcal M \subset \mathbb C^N$. A multidimensional generalization of Cauchy's theorem of complex analysis guarantees, under some conditions on $\mathcal M$, that the integral over $\mathcal M$ and $\mathbb R^N$ of any holomorphic integrand is the same.   This allows one to compute expectation values of observables $\mathcal O$ for which $\mathcal O e^{-S}$ is holomorphic, if $\mathcal M$   is properly chosen. The key to these methods is that the average sign $\left< e^{-i S_I}\right>$ is an integral of a non-holomorphic function, and therefore dependent upon the integration manifold, whereas the physical expectation values do not. The manifold $\mathcal M $ was originally suggested to be the combination of thimbles, multidimensional generalizations of the steepest descent/constant phase path familiar from complex analysis. This method and its associated algorithmic problems were pursued by several groups~\cite{Cristoforetti:2014gsa,DiRenzo:2015foa,Mukherjee:2013aga,Fujii:2015vha,Fukushima:2015qza,Alexandru:2015xva,Alexandru:2016lsn,Alexandru:2015xva,Alexandru:2016gsd,Alexandru:2015sua,Alexandru:2016san,Alexandru:2017czx,Alexandru:2016ejd,Alexandru:2017lqr,Alexandru:2017oyw}. Relevant analytical work, closely connected to the ``resurgent transseries" (for a recent review see~\cite{Aniceto:2018bis}) was also pursued in~\cite{Tanizaki:2016xcu,Tanizaki:2016cou,Tanizaki:2015rda,Kanazawa:2014qma,Schmidt:2017gvu}. 
Experience with actual simulations made evident some problems with the thimble approach.
The first is that thimbles are complicated manifolds that have to be found ``on the fly" by the algorithm and the lack of a local characterization of thimbles makes this computationally expensive. Second, theories where more than one thimble contribute to the path integral  significantly~\cite{Tanizaki:2016cou}  are particularly difficult to sample~\cite{Alexandru:2015xva,Alexandru:2016san}. 

This led to some modifications of the method. In the
generalized thimble method~\cite{Alexandru:2015sua,Alexandru:2016ejd,Alexandru:2017lqr} 
the manifold of integration $\mathcal M$ is chosen to be the deformation of $\mathbb R^N$ by the holomorphic flow defined by the action. If $\mathbb R^N$ is deformed by the flow by an infinite amount of flow time, $\mathcal M$ approaches the right combination of thimbles equivalent to the original integration domain. If the flow is stopped at some finite flow time, $\mathcal M$ is close, but not identical, to the sum of appropriate thimbles. It is, however, a legitimate manifold of integration in the sense that it gives exactly the same result as the original manifold $\mathbb R^N$. The advantage of the manifold $\mathcal M$ over the thimbles is that 1) less flow corresponds to smaller computational cost and 2) $\mathcal M$ can be algorithmically constructed during the simulation by solving the flow equations while finding the thimbles and determining which ones contribute to the integral is a difficult task in all but the simplest field theories. 

This is not to say that the generalized method does not have its own problems.
 Large flow times can improve the sign problem  but generate multimodal distributions difficult to sample. Shorter flowing times avoid the multimodality but improve the sign problem less, so the flow time has to be carefully chosen and, in fact, there is no guarantee that a ``middle ground" flowing time can be found. (Multimodality can also be dealt with by more sophisticated sampling algorithms~\cite{Fukuma:2017fjq,Alexandru:2017oyw}.)  In addition, the computation of the Jacobian arising from parameterizing $\mathcal M$  by the initial point of the flow in $\mathbb R^N$ is expensive. The proposal presented in this paper drastically reduces the cost of the Jacobian. At the same time, it provides more flexibility in the choice of $\mathcal M$ while systematically improving the sign problem.

 One step towards speeding up the costly calculations involved in the generalized thimble method was given in~\cite{Alexandru:2017czx}. A feed-forward neural network was trained to interpolate points in $\mathcal M$ obtained by the more expensive holomorphic flow. The neural net was then used to quickly generate more points in $\mathcal M$.
In the present paper we go one step further and completely bypass the need to generate points by flowing. Instead, we seek to flow directly toward a manifold of maximum average sign, albeit in a restricted family of manifolds $\mathcal M_\lambda$ which are parameterized by a finite number of parameters $\lambda$.  A similar proposal based on maximizing the approximate average sign was pursued in~\cite{Mori:2017nwj,Mori:2017pne}.  Our method produces a manifold $\mathcal M$ that can be sampled as rapidly as $\mathbb R^N$ via :
\begin{widetext}
\begin{equation}\label{eq:expectation-parameterized}
\left<\mathcal O\right>
=\frac{\int_{\mathcal M}\mathcal D\tilde\phi\;\mathcal O(\tilde\phi) e^{-S(\tilde\phi)}}{\int_{\mathcal M}\mathcal D\tilde\phi\; e^{-S(\tilde\phi)}}
=\frac{\int_{\mathbb R^N}\mathcal D\phi\;\mathcal O[\tilde\phi(\phi)] e^{-S[\tilde\phi(\phi)]} \det J(\phi)}
{\int_{\mathbb R^N}\mathcal D\phi\; e^{-S[\tilde\phi(\phi)]} \det J(\phi)}
\equiv
\frac
{\int_{\mathbb R^N}\mathcal D\phi\;\mathcal O[\tilde\phi(\phi)] e^{-\Seff(\phi)}}
{\int_{\mathbb R^N}\mathcal D\phi\; e^{-\Seff(\phi)}},
\end{equation}
\end{widetext} 
where a point $\tilde\phi$ in $\mathcal M$  is parametrized by a point $\phi$ in $\mathbb R^N$. 
 We present in Sec.~\ref{sec:algorithm} how the algorithm can be implemented, with special emphasis on the gradient ascent method we use to obtain the local maximum value of average sign.  Further, it is shown that the derivative of the sign problem with respect to $\lambda$ can be efficiently calculated despite a potentially small sign.

 The method of determining an optimal manifold for integration, as well as the procedure for integrating along that manifold, is detailed in Sec.~\ref{sec:algorithm}. In Sec.~\ref{sec:model}, we define the physical model we study with this algorithm, the Thirring model. In Sec.~\ref{sec:results} we present our results, and conclusions are summarized in Sec.~\ref{sec:discussion}.


\section{The Method}\label{sec:algorithm}

We start by specifying a family $\mathcal M_\lambda$ of submanifolds of $\mathbb C^N$, parameterized by $\lambda$. The choice of this family is guided by the ease of computation of the Jacobian and some experience
acquired with the generalized thimble method.
We then 
proceed to maximize the average sign among this family of manifolds using
 a simple gradient ascent technique.  On a manifold of integration $\mathcal M_\lambda$, the average sign is
\begin{equation}\label{eq:sign1}
\left<\sigma\right>_{\lambda}
=
\frac
{\int_{\mathbb R^N} \mathcal D\phi\;e^{-\Seff[\phi;\lambda]}}
{\int_{\mathbb R^N} \mathcal D\phi\;e^{-\Re\Seff[\phi;\lambda]}},
\end{equation} 
where $\phi$ are the fields in the theory and $S_\text{eff}\equiv S-\ln\det J$ is the effective action.
On this manifold, we compute a vector proportional to the gradient of the magnitude of the average sign, and then proceed to change $\lambda$ by a small amount along this vector.
\begin{equation}
\lambda_s - \lambda_{s-1} \propto \eta
\nabla_\lambda |\left<\sigma\right>_{\lambda}|
\end{equation}
Here $\eta$ is the learning rate, determining how large each step along the computed gradient should be.  We initialize $\mathcal M_{\lambda_0}$ to be $\mathbb R^N$.   After a large number of steps, and if the learning rate $\eta$ is small enough, we should arrive at a (local) maximum of the average sign.  Critically, the computation of the direction of the gradient has no sign problem.

We now show how to compute the direction of the gradient. The numerator of Eq.~(\ref{eq:sign1}), being the integral of a holomorphic function $e^{-S}$ along $\mathcal M$, does not depend on $\lambda$.  In contrast, since the integral of $e^{-\Re \Seff}$ cannot be written as an integral of a holomorphic function, the denominator will vary with $\lambda$. The gradient of the magnitude $\left|\left<\sigma\right>_\lambda\right|$ with respect to the manifold parameters $\lambda$, then, is given by 
\begin{widetext}
\begin{align}
\nabla_\lambda \left|\left<\sigma\right>_{\lambda}\right|
&= -\left|\left<\sigma\right>_{\lambda}\right| \ 
\frac
{\nabla_\lambda \int_{\mathbb R^N} \mathcal D\phi\;e^{-\Re \Seff[\phi;\lambda]}}
{\int_{\mathbb R^N} \mathcal D\phi\;e^{-\Re \Seff[\phi;\lambda]}}\nn\\
&= \left|\left<\sigma\right>_{\lambda}\right| \ 
\frac
{\int_{\mathbb R^N} \mathcal D\phi\;e^{-\Re \Seff[\phi;\lambda]}\left[\nabla_\lambda S_R - \Re \Tr J^{-1} \nabla_\lambda J\right]}
{\int_{\mathbb R^N} \mathcal D\phi\;e^{-\Re \Seff[\phi;\lambda]}}\,.
\end{align}                                                                                                                                                                                                                                                                                                                                                                                                                                                                        \end{widetext} 
From this, we see that the gradient factorizes into two pieces: the average sign on $\mathcal M_\lambda$, and an expectation value of an operator on that manifold.
	The second factor is an expectation value with respect to $e^{-\Re\Seff}$, and therefore is sign-problem free; the first is a scalar which does not affect the direction.  This allows us to compute (up to that overall scalar) the gradient on a manifold reliably by a short Monte Carlo simulation. For a gradient ascent method, an overall magnitude like $\left|\left<\sigma\right>_{\lambda}\right|$ (even varying with $\lambda$) can be safely neglected: it does not change the direction the gradient points in $\lambda$-space.  This allows our method to be efficient even when the average sign is statistically indistinguishable from zero. Therefore, at each step $s$ of the gradient ascent, we update the manifold parameters $\lambda_s$ according to
\begin{equation}\label{eq:grad}
\lambda_s - \lambda_{s-1} = \eta
\left<
\nabla_\lambda S_R - \Re \Tr  J^{-1} \nabla_\lambda J
\right>_{\Re\Seff}
\end{equation}
In principle, one might use a more efficient stochastic gradient ascent algorithm, such as \textsc{Adam}~\cite{2014arXiv1412.6980K}, to both speed up the calculation and avoid finding suboptimal local maximum. For this work, we found na\"ive gradient ascent converges adequately swiftly, and the stochastic nature of the Monte Carlo simulation used to compute the gradient helped to explore parameter space.

In a gradient ascent method, the parameter $\eta$ must chosen to be small enough to avoid overshooting a maximum, but not much smaller, otherwise it will oscillate around the maximum but never converge. For our purposes, there is one additional practical consideration restricting the size of $\eta$. In calculating the expectation value of Eq.~(\ref{eq:grad}), we would like to avoid needing to completely re-thermalize the Markov chain after every gradient ascent step. To this end, we set the step size $\eta$ to be sufficiently small that $S[\tilde\phi(\phi)]$, for any fixed $\phi$, changes only slowly with $s$. The value of $\phi$ at the end of one Monte Carlo run can then be used to seed the next run on the new manifold, minimizing the necessary thermalization time.
	
It should be stressed that lack of care in this process, or in any other detail of the sign maximization process, may reduce the average sign of the manifold ultimately found and increase the computational time, but does not affect the correctness of physical observables on that manifold. The ``real-plane'' integral is calculated as an integral over compact variables, that is, an integral over $\mathbb T^N = (S^1)^N$. The manifold $\mathcal M_\lambda$ is a submanifold of the complexified $N$-torus $(S^1 \times \mathbb R)^N$. Cauchy's integral theorem guarantees that, provided the domain of integration is compact (as $\mathbb T^N$ is), the integral over $\mathcal M_\lambda$ will equal that over $\mathbb T^N$ if the manifold $\mathbb T^N$ is continuously deformable to $\mathcal M_\lambda$. For our purposes, this is guaranteed by making the family $\mathcal M_\lambda$ be continuous in the parameters $\lambda$, and letting $\mathcal M_\lambda = \mathbb T^N$.

The determinant of the Jacobian $J$ -- which must be computed during a Monte Carlo on $\mathcal M$ -- is a potentially expensive operation, with a cost approximately cubic in the number of degrees of freedom. To avoid this, we will chose an ansatz family $\mathcal M_\lambda$ for which the Jacobian is diagonal. In particular, we write $\tilde\phi_i(\phi) = \phi_i + i f_i(\phi_i)$, so that $J_{ij} = \delta_{ij}\left(1 + f'_i(\phi_i)\right)$, which is the most general ansatz possible satisfying our constraints.  
Relaxing this constraint to a non-diagonal Jacobian should improve the sign problem by allowing nonlocal correlations in the imaginary components of $\phi$, but this will come at computational expense and will be left to future work.

\section{Thirring model}\label{sec:model}
In order to make the ideas more concrete we will phrase our discussion in terms of a specific field theory model,  the $1+1D$ massive Thirring model with Wilson fermions.
The lattice action is given by
\begin{equation}\label{eq:lattice-action}
S=\sum_{x,\nu} \frac{N_F}{g^2} (1-\cos A_\nu(x))+\sum_{x,y} \bar\psi^a(x) D^{W}_{xy}(A)  \psi^a(y)
\end{equation}
with
\begin{equation}
D^W_{xy} = \delta_{xy} - \kappa \sum_{\nu=0,1}  
\Big[ 
 (1-\gamma_\nu) e^{i A_\nu(x)+\mu \delta_{\nu 0}} \delta_{x+\nu, y}
 + (1+\gamma_\nu) e^{-i A_\nu(x)-\mu \delta_{\nu 0}}  \delta_{x, y+\nu}
\Big], \nn
\end{equation} where $\psi$ is a two-component Dirac spinor with the flavor indices $a$ taking values from $1,\ldots,N_F$, $g$ is the coupling, $\mu$ the fermion chemical potential and $\kappa=1/(2m+4)$, where $m$ is the bare mass of the fermions.  Standard universality arguments applied to this asymptotically free theory indicate that, in the continuum limit, this action is equivalent to the continuum action

\begin{equation}
S = \int \d^2 x\left[
\bar\psi^a(\slashed\partial + \mu\gamma_0+m)\psi^a + \frac{g^2}{2N_F}\bar\psi^a\gamma_\mu\psi^a\bar\psi^b\gamma_\mu\psi^b
\right]\text.
\end{equation}
The four-fermion interaction is generated when the bosonic $A_\mu(x)$ auxiliary field is integrated over. The Thirring model was chosen since other similar methods have been applied to it, thus it serves as a useful benchmark for our method.

%
The integration over the fermion fields results in the action 

\begin{equation}\label{eq:action}
S_{\rm eff}=N_F 
\left(  
\frac{1}{g^2}\sum_{x,\nu} (1-\cos A_\nu(x)) - \log\det D(A)
\right)
\text.
\end{equation}
In this work we take $N_F = 2$. For finite chemical potential $\mu \ne 0$, the determinant $\det D(A)$ is not strictly real, and we must address a sign problem.

In applying the method described in Sec.~\ref{sec:algorithm} we enforce
 three additional constraints on $f$, all coming from symmetries of the action Eq.~(\ref{eq:action}). The action is $2\pi$-periodic in the fields $A_0,  A_1$ and an even function; therefore we require the same of $f$. Finally, the action is invariant under translations of the lattice. The lattice degrees of freedom are divided into timelike links $A_0$ and spatial links $A_1$. Translational invariance of the $f_i$ implies that the form of $f_i$ can depend only on whether the index $i$ refers to an $A_0$ field or an $A_1$ field.

Consistent with these demands, we use a simple two-parameter family with $f_0(\phi) = \lambda_0 + \lambda_1 \cos\phi$ and $f_1(\phi) = 0$, so that the manifold $\mathcal M_\lambda$ is defined by
\begin{align*}
\tilde A_0(A_0, A_1) &= A_0 + i\left(\lambda_0 + \lambda_1 \cos A_0\right)\\
\tilde A_1(A_0, A_1) &= A_1
\end{align*}

As discussed above, Cauchy's theorem guarantees that expectation values computed on $\mathcal M_\lambda$ are equal to those computed on $\mathbb T^N$, provided that one manifold may be continuously deformed to the other. To see that this is so, note that $\mathcal M_0 = \mathbb T^N$, and that $\tilde A$ is a continuous function of $\lambda$.
One might consider using a larger class of manifolds.  We have investigated including a $\cos^2 A_0$ term in $\tilde A_0(A_0, A_1)$ and a $\cos A_1$ in the $\tilde A_1(A_0, A_1)$, but in all cases found negligible improvement in the average sign computed on the resulting manifolds.   

Once the manifold has been selected by a suitably long gradient ascent, we perform a Monte Carlo calculation to determine observables of interest via Eq.~(\ref{eq:expectation-parameterized}). The imaginary part of both the action and the log of the Jacobian determinant must be included in the reweighting. 
Since we chose a manifold of integration for which the Jacobian is diagonal, the Monte Carlo sampling proceeds as quickly as it would for a standard Metropolis running on $\mathbb R^N$. There are no constraints on the observables computed, aside from the requirement that $\mathcal O e^{-S}$ be holomorphic.

\section{Results}\label{sec:results}
We choose bare parameters $g$ and $m$ of the action so that the renormalized particle masses lie below the lattice cutoff scale. We measure two particle masses -- a fermion mass $a m_f$ and a boson mass $a m_b$ -- by fitting the large-time behavior of correlators $\left<\mathcal O_\alpha(t) \mathcal O_\alpha(0)^\dagger\right>$, where $\mathcal O_f = \psi_1$ and $\mathcal O_b = \bar\psi_i\gamma_5(\tau_3)_{ij}\psi_j$ (the fermion subscripts denote flavor). For simulations in this paper, we take $g = 1.0$ and $m=-0.25$, leading to renormalized masses of $am_f = 0.30(1)$ and $am_b = 0.44(1)$. We then have $m_b / m_f = 1.5(2)$, corresponding to a strongly coupled theory since the binding energy of the boson is comparable to the rest mass of the constituent fermions.

\begin{figure}[t]
 \includegraphics[width=0.48\linewidth]{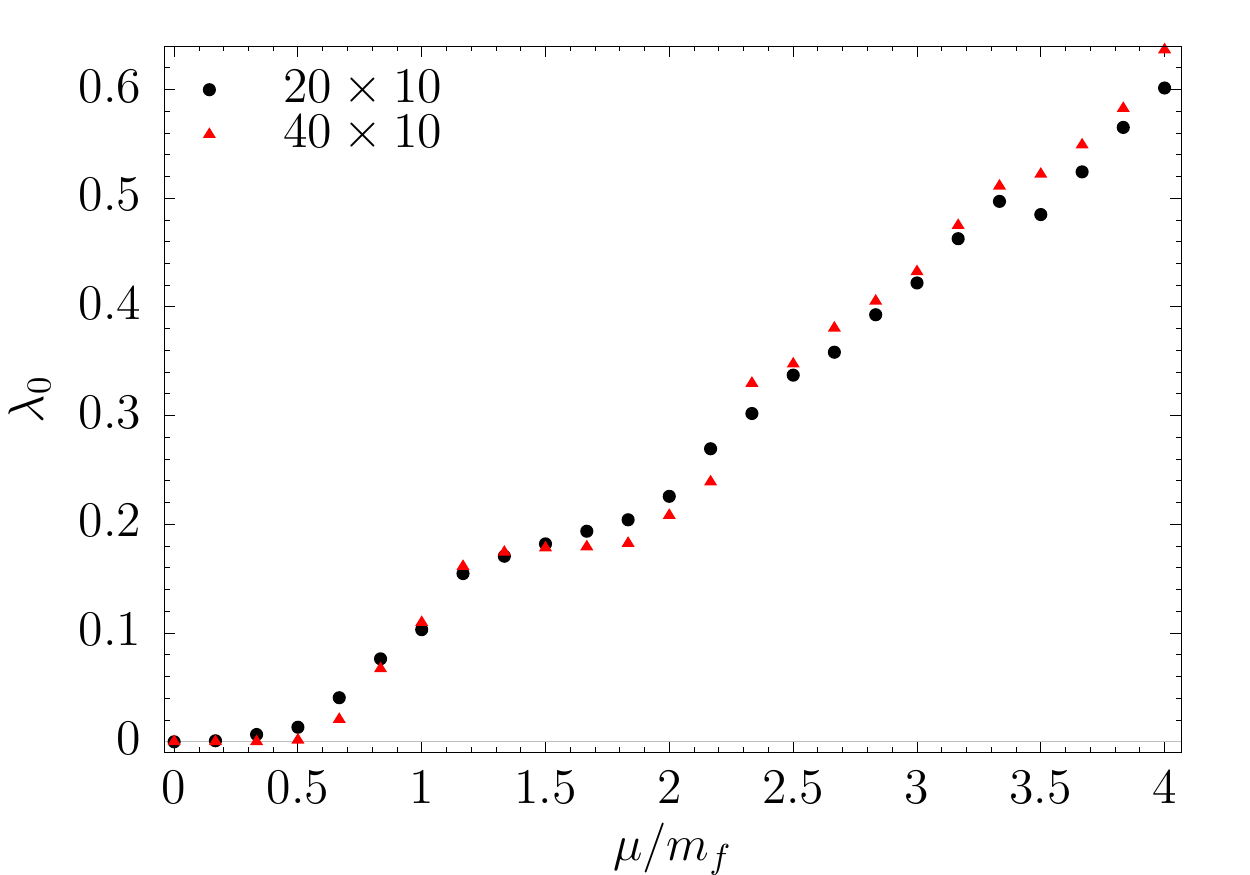}
 \includegraphics[width=0.48\linewidth]{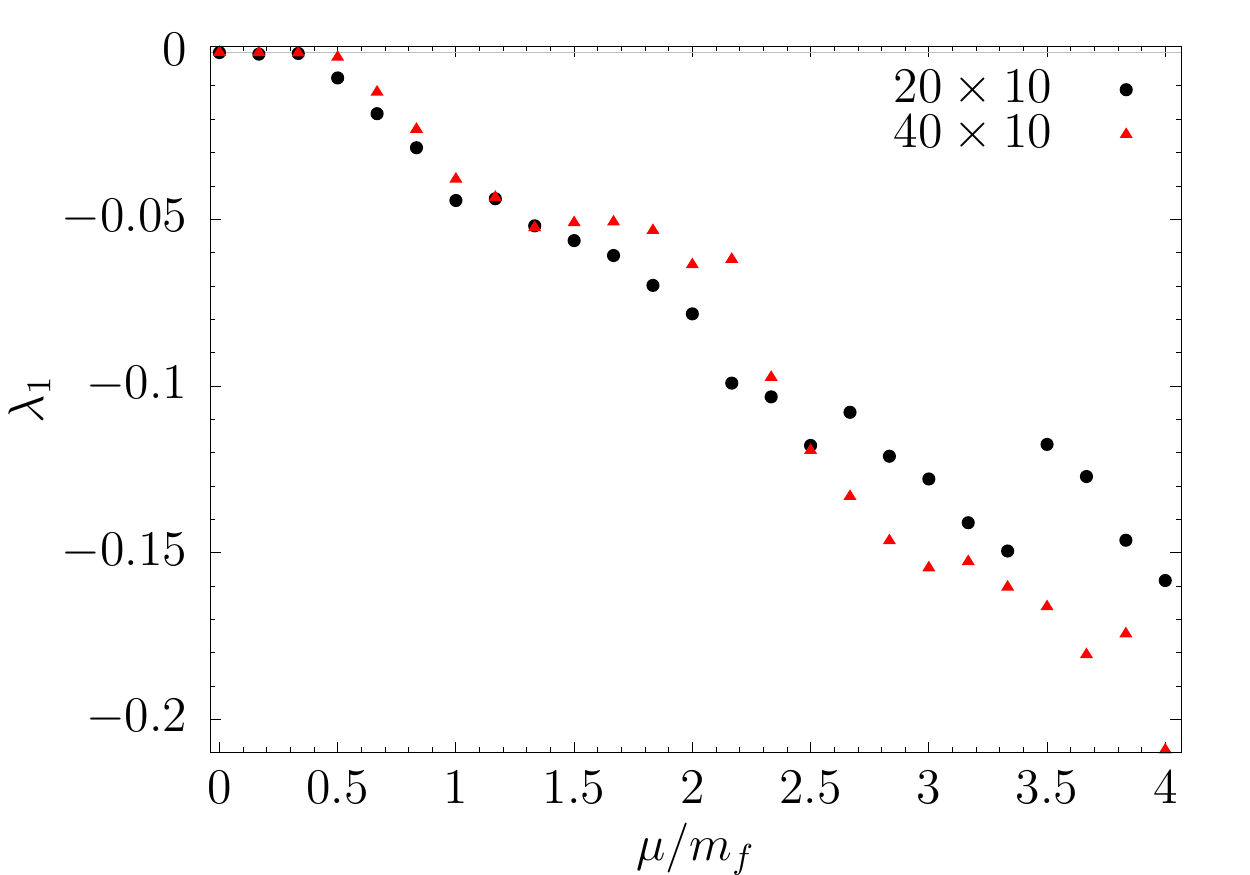}
\caption{Manifold parameters as a function of chemical potential $\mu/m_f$, for both 20$\times$10 and 40$\times$10 lattices, with bare parameters $m=-0.25$ and $g=1$.\label{plt:manifold}}
\end{figure}
We perform calculations on two lattice sizes: $N_t\times N_x=20\times10$ and $40\times10$.  
The maximization of the sign average is done using a step size $\eta=10^{-4}$. 
This step size was determined by starting with a large $\eta$, 
where the optimization process exhibited oscillatory behavior, and then 
reducing it until the process becomes smooth.
We only tuned it on the most demanding ensemble, the $40\times 10$ ensemble with the largest chemical potential, and used the same value for all other ensembles.
The optimization process is stopped when the $\lambda$ parameters converge, that is when
the gradient in Eq.~(\ref{eq:grad}) becomes too small.
The manifold parameters determined by the optimization procedure are shown in Fig.~\ref{plt:manifold}. For both lattice sizes, the chosen parameters appear to be nearly continuous functions of $\mu/m_f$. This suggests a simple optimization going forward: perform the gradient ascent at a small number of values of $\mu$, and interpolate to determine the manifold of integration for all other desired chemical potentials. Another option is to use as a starting point for the optimization process the values determined for a ``nearby'' ensemble, one with similar chemical potential. 

The fact that the interpolated values of $\lambda_0, \lambda_1$ are approximate and not strictly optimal affects only the efficiency of the algorithm, not its correctness. For more elaborate families of manifolds, with more parameters, the gradient ascent phase becomes more time-consuming. This optimization could be computationally expensive in such cases, but optimizations along the lines suggested above are likely to be available. We note that the discontinuities in Fig.~\ref{plt:manifold} are due to an early exit from the optimization loop. We decided to keep these parameters to show that this discontinuity
is not reflected in the observables, as a further check of the method.

\begin{figure}[t]
 \includegraphics[width=0.48\linewidth]{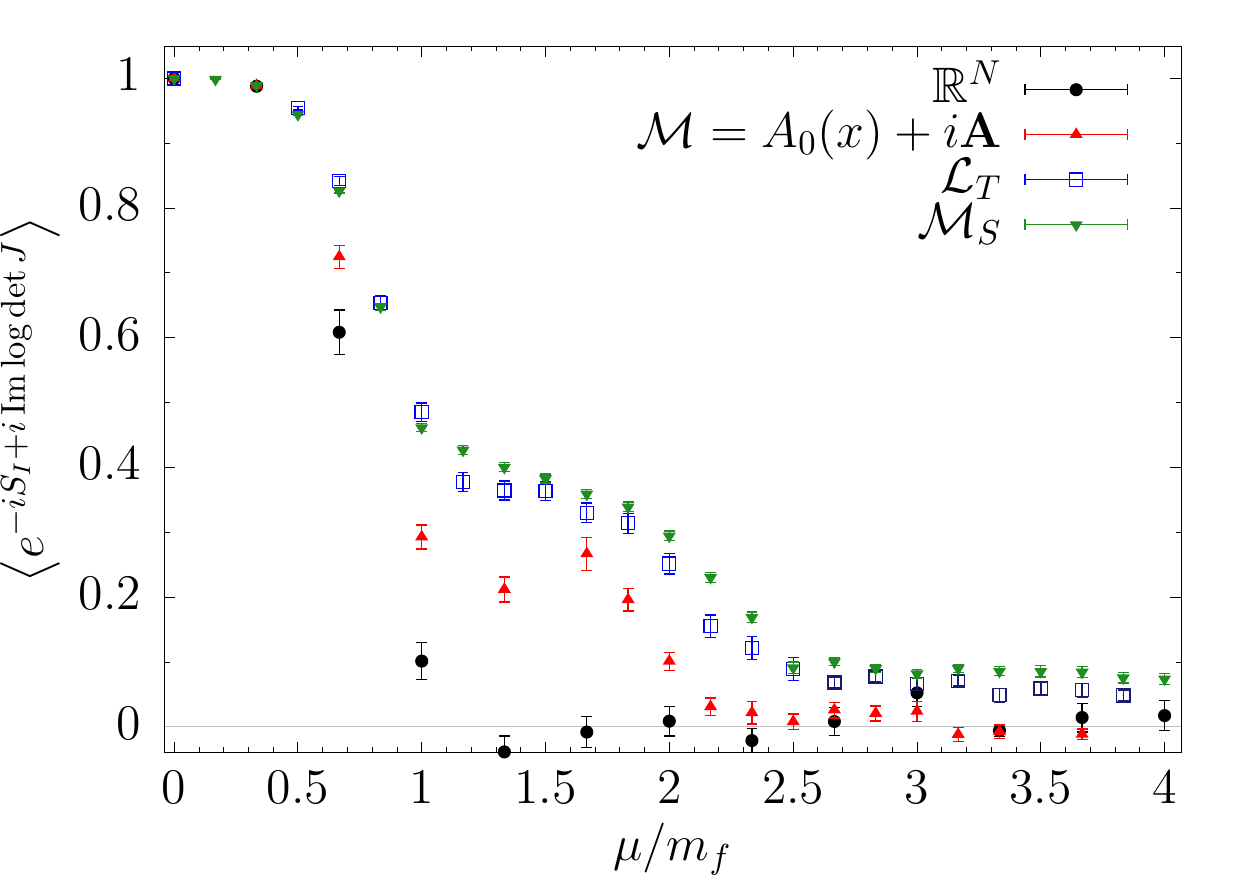}
 \includegraphics[width=0.48\linewidth]{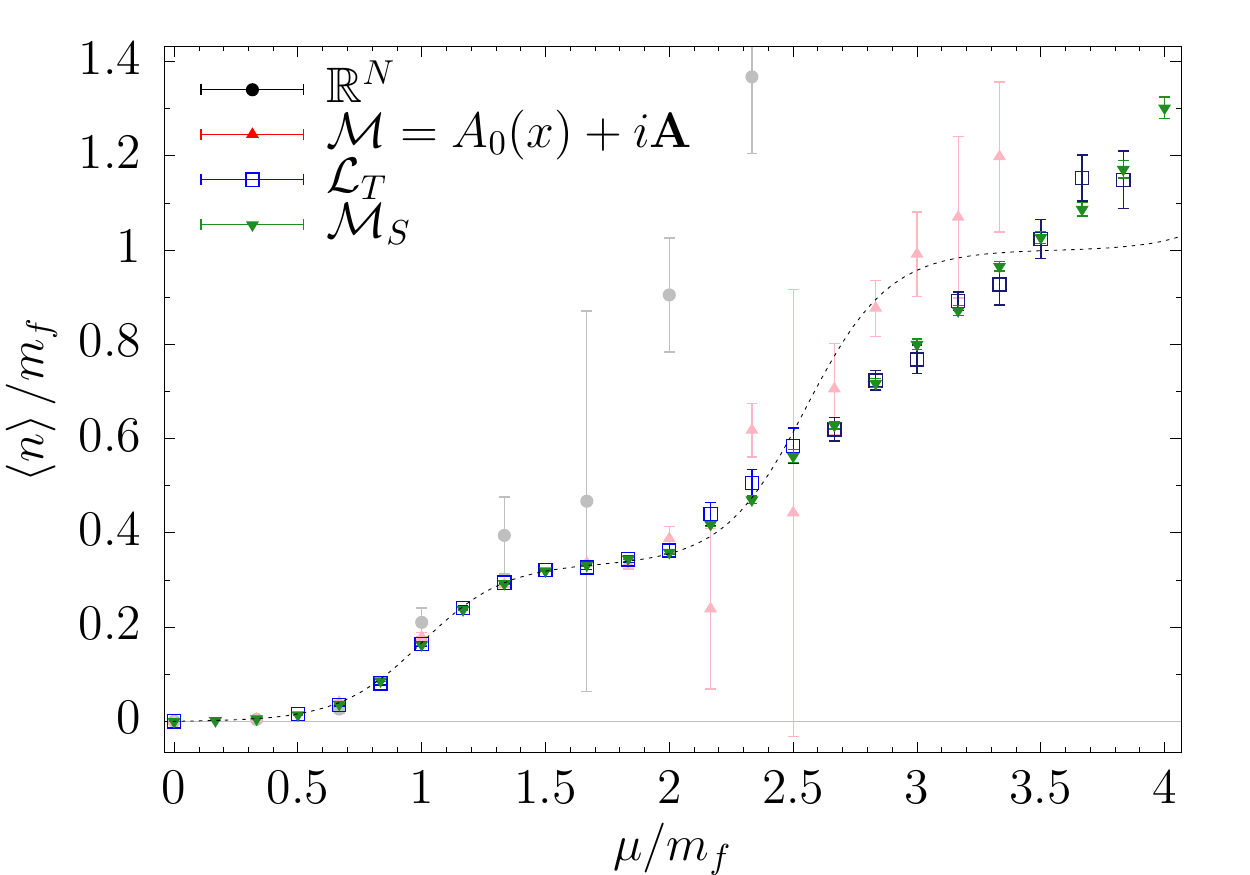}
\caption{$\langle e^{-iS_I+i\im\log\det J}\rangle$ and $\langle n\rangle/m_f$ as a function of $\mu/m_f$ for Wilson fermions on lattices of size $20\times10$.  The dashed curve represents the free fermion gas with the same mass.\label{plt:20}}
\end{figure}

With the parameters determined above, we performed a Monte-Carlo calculation generating
of the order of two to ten thousand independent configurations (except for a few points discussed
below).
The average sign and measurements of average fermion density (per flavor) $\left<n\right>$ for $0 < \mu/m_f < 4$ on a 20$\times$10 lattice are shown in Fig.~\ref{plt:20}. The real plane ($\mathbb R^N$) calculations are shown in black; data points for which the average sign could not be distinguished from $0$ at $2\sigma$ (indicating that no measured observable will be meaningful) are grayed out.  Calculations on the tangent plane of the dominant Lefschetz thimble $A_0(x)+i\textbf{A}$ are shown in red, and those of the machine-learned \textit{learnifold} $\mathcal L_T$ from Ref.~\cite{Alexandru:2017czx} in blue. Finally, we present calculations done on the sign-optimized manifold $\mathcal M_S$ in green. We see that the sign-optimized manifold finds an average sign problem as good or better than the learnifold does, with the added benefits of being computationally faster and simpler to implement.  These improvements allow for us to compute the density with reduced uncertainty and even reach higher values of $\mu/m_f$.
As a further check of our results, we show the result for non-interacting fermions (with the same renormalized mass) as a dotted line.

\begin{figure}[t]
 \includegraphics[width=0.48\linewidth]{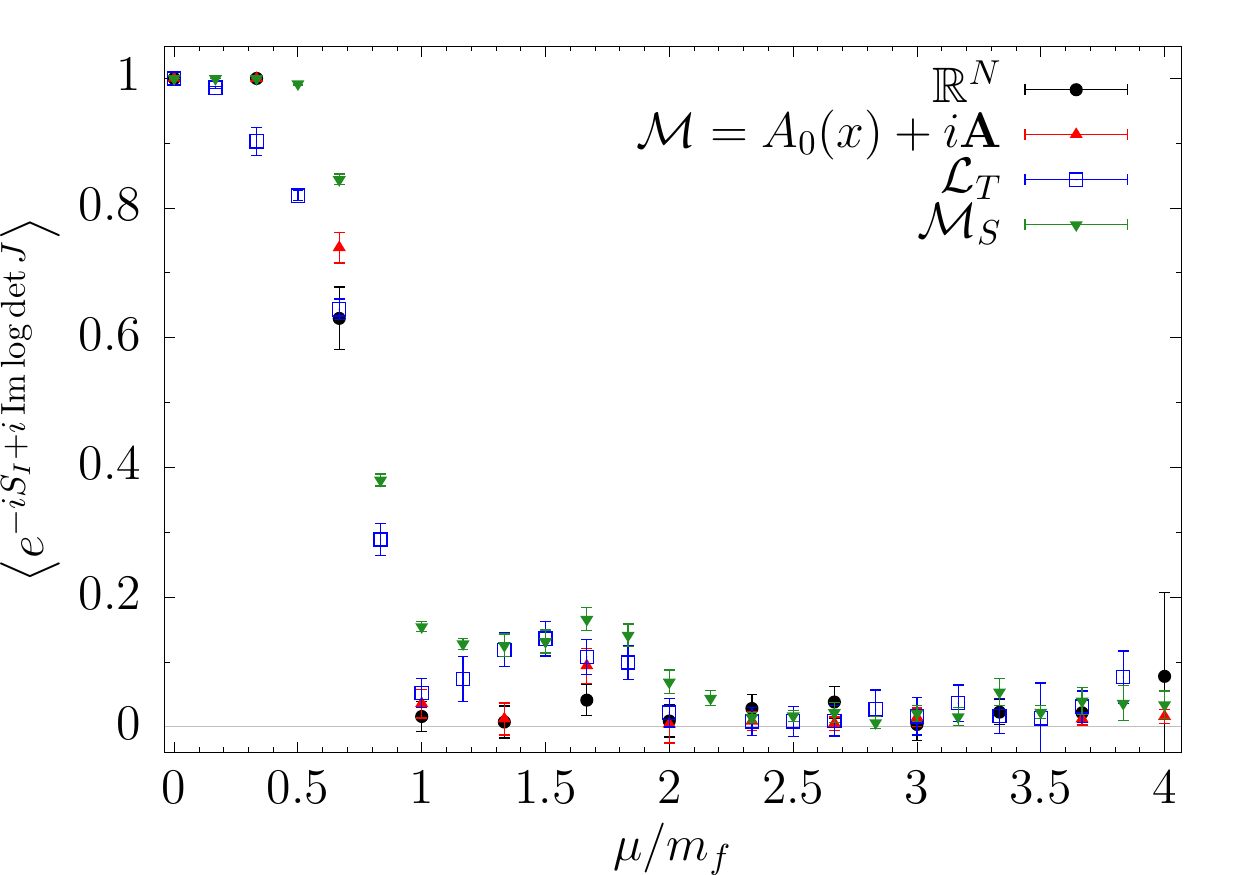}
 \includegraphics[width=0.48\linewidth]{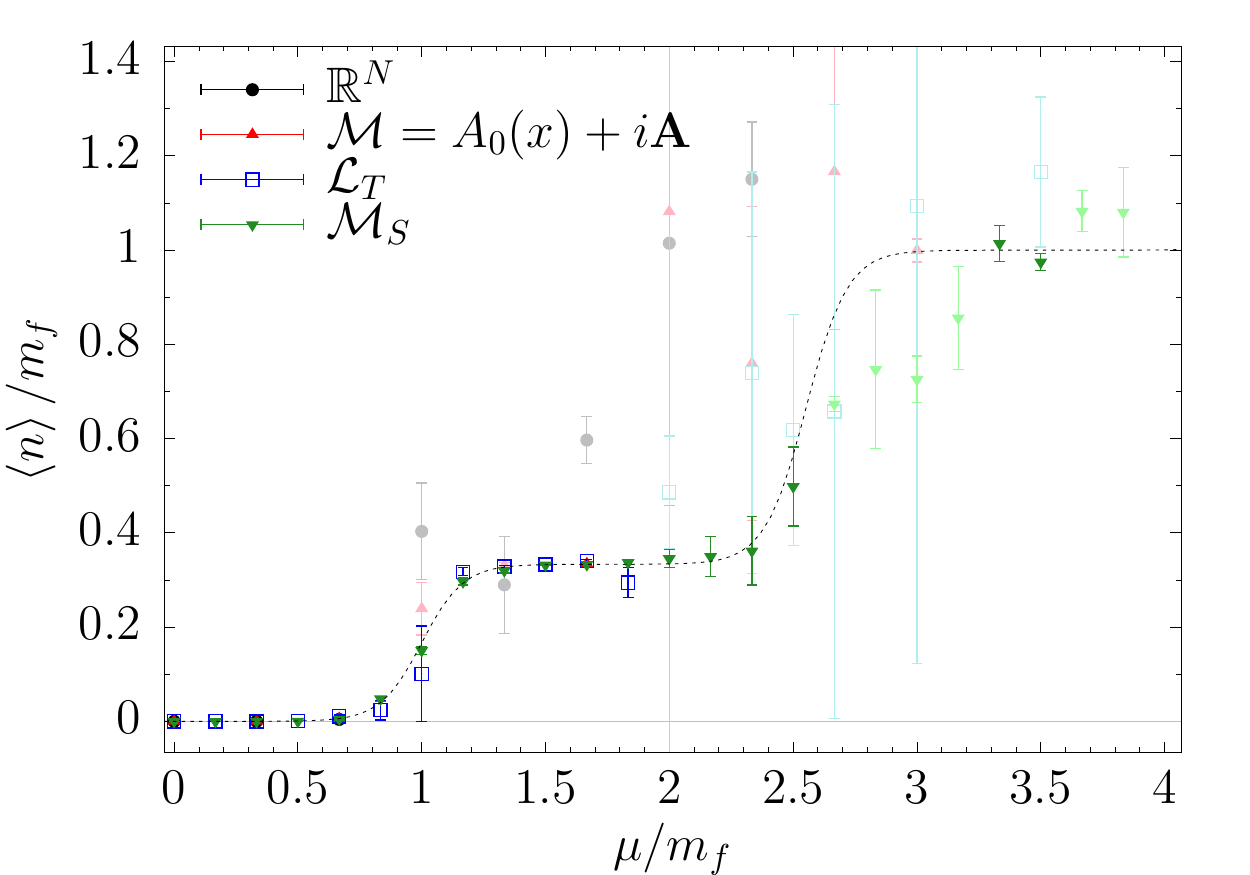}
\caption{$\langle e^{-iS_I+i\im\log\det J}\rangle$ and $\langle n\rangle/m_f$ as a function of $\mu/m_f$ for Wilson fermions on lattices of size $40\times10$.  The dashed curve represents the free fermion gas with the same mass.\label{plt:40}}
\end{figure}

Similarly, results for a lattice size of $40\times10$ are shown in Fig.~\ref{plt:40}. On this larger lattice, the relative performance of the sign-optimized manifold is moderately improved. For $\mu/m_f > 1.83$, neither the real plane calculation, nor the learnifold, could resolve the sign problem.  The sign-optimized manifold has sufficiently large average signs to allow us to measure the density up to $\mu/m_f = 2.50$.  Furthermore, other methods had problems computing the density near $\mu/m_f \approx 1.00$, while using the sign-optimized manifold we compute the density at this point easily.

At both lattice sizes, we demonstrate that the sign-optimized manifold method is capable of reproducing the ``Silver Blaze'' phenomenon \cite{Cohen:2003kd}: the $\mu$-independence of observables below the threshold chemical potential $\mu \approx m_f$. 

To estimate the speedup given by the optimized manifold over a naive calculation on $\mathbb R^N$, we performed two tests.  First, $72,000$ decorrelated measurements at $\mu/m_f=3.33$ on a $20\times10$ lattice. This number of measurements is not enough to resolve the sign average from zero, so we obtain only a lower bound on the speedup attributable to using $\mathcal M_S$. We find that the real plane has an average sign of $0.002 \pm 0.003$, whereas the $\mathcal M_S$ has an average sign of $0.086 \pm 0.007$, which is larger by at least a factor of $16$.  The number of measurements required to obtain a fixed precision is proportional $\left<\sigma\right>^{-2}$, therefore this corresponds to a speedup greater than $250$.  The second test computed $10,000$ measurements on $\mathbb R^N$ at $\mu/m_f=1.00$ on a $40\times10$ lattice (this value is of interest because it corresponds to the first particle threshold).  The real plane was found to have an average sign of $0.005\pm0.005$, but an average sign of $0.155\pm0.007$ on the optimized manifold, which is larger by at least a factor of 15, giving a speedup of $225$.

The speedup given by this algorithm over the learnifold procedure is more difficult to estimate; however, the learnifold procedure requires evolving the holomorphic flow equations many times to achieve at best the same average sign as $\mathcal M_S$.  According to Ref.~\cite{Alexandru:2017czx}, generating the training set and training the neural network took 114 CPU-hours for a $20\times10$ lattice at $\mu/m_f=3.83$.  The algorithm described in this paper replaces that step with a gradient ascent routine, which took approximately 24 CPU-hours, and is amenable to further optimization.

\section{Discussion and prospects}\label{sec:discussion}
We have exhibited an efficient method for reducing the sign problem of the finite density Thirring model in 1+1 dimensions. Our method works with a pre-determined family of manifolds, seeking the manifold in that family which has the largest average sign. Once such a manifold has been found, a standard Metropolis calculation, with reweighting, is performed on that manifold.  Using this method, we have increased the $\mu/m_f$ range that can reliably computed. It is important to stress that comparisons with other methods of dealing with the sign problem must take into account that the computational cost of the method presented here has both a fixed cost (independent of the number of measurements made) and a variable one (that is proportional to the number of measurements). The variable cost compares very favorably with other methods, especially the generalized thimble method. Therefore, the best way to apply the method is to determine the parameters of $\mathcal M$ roughly so the average sign is distinguishable from zero but not necessarily particularly close to one. Then, a high number of measurements can be made cheaply to reduce the error bars. 

This method is closely related to previous approaches based on the complexification of lattice degrees of freedom, but works without evolving a differential equation to determine the manifold of integration. This makes it faster as long as the parameters defining the manifold can be determined quickly. The method has the drawback that it requires the construction of a model-specific family of manifolds, so physical insight is required. Nevertheless, given such an ansatz, the method is very advantageous. 
This suggests that theoretical effort should be put into generating ansatze applicable to more interesting physical theories, like
 gauge theories and real-time (Minkowski space) calculations of other models.

\begin{acknowledgments}
A.A. is supported in part by the National Science Foundation CAREER grant PHY-1151648 and by U.S. Department of Energy grant DE-FG02-95ER40907. A.A. gratefully acknowledges the hospitality of the Physics Departments at the Universities of Maryland and Kentucky, and the Albert Einstein Center at the University of Bern where part of this work was carried out. 
P.F.B., H.L., and S.L.  are supported by U.S. Department of Energy under Contract No. DE-FG02-93ER-40762.
\end{acknowledgments}
\bibliographystyle{apsrev4-1}
\bibliography{thimbology}
\end{document}